\pdfoutput=1 

\RequirePackage{lineno}
\documentclass[cits] {JINST}
  \usepackage{multicol}
  \usepackage{float}
  \usepackage[numbers,nonamebreak]{natbib}
\title{Energy and Direction Estimation of Neutrinos in muonless events at ICAL}

\author{Ali Ajmi$^a$,
and S. Uma Sankar$^b$\\
\llap{$^a$}Homi Bhabha National Institute,\\
  Anushaktinagar, Mumbai 400 094, India\\
\llap{$^b$}Indian Institute of Technology Bombay,\\
  Mumbai 400 076, India\\
E-mail: \email{aliajmi@tifr.res.in},\\
  \email{uma@phy.iitb.ac.in}}

  \abstract{
  In this paper, we study events without identifiable muon tracks in the Iron Calorimeter detector at the India-based Neutrino Observatory. Such events are dominated by high energy (E$_\nu>$1 GeV) $\nu_e$ charged current interactions, which have been studied only in a few experiments so far. The charged particles, produced in these neutrino interactions, give rise to a set of hits in the detector. We attempt to 
reconstruct the energy and the direction of the neutrino in such events. We study the energy distribution for a given 
pattern of hits of these events and find that the Landau distribution provides a good fit. 
We define two kinematic variables based on the hit distribution and use
them to determine the cosine of the polar angle of the neutrino direction ($\cos \theta$). There is 
a moderate correlation between these variables and the $\cos \theta$. These provide us enough information to prepare calibration charts for looking up the energy and direction of the incident neutrino.
}

\begin{document}


\section{Introduction}
 India-based Neutrino Observatory (INO) \cite{Athar:2006yb} is an upcoming experimental facility which aims to study atmospheric neutrinos. 
It houses a gigantic magnetised Iron Calorimeter (ICAL) neutrino detector. ICAL consists of 3 modules, each of which contains 151 horizontal iron layers. These iron layers, each of thickness 5.6 cm, are interspersed with resistive
plate chambers (RPCs) \cite{Cardarelli:2007eb,Riegler:2002vg,Lippmann:2003uaa}. The total number of RPCs in the detector is $\sim$30,000, and the total mass of the detector is approximately 50 kilotons. 
When a neutrino interacts with an iron nucleus, it produces
a set of charged particles. Depending on the particle type and its energy, these charged particles 
pass through one or more RPCs. Whenever a charged particle passes through an
RPC, it produces a hit. 
These hits are our primary signals. The layer number of RPC gives the z-coordinate of the hit. The x and y-coordinates are
given by the copper-strips of the pick-up panels which are orthogonally oriented at the top and 
the bottom of the RPCs  \cite{2008arXiv0810.4693B}. 

The ICAL detector at INO will observe different types of neutrino interactions. The charged current (CC) interactions
of the muon neutrinos ($\nu_\mu$ and $\bar{\nu}_\mu$) produce events with muons ($\nu_\mu$CC events). 
If the energy of the muon is $>$0.5 GeV, and its direction not close to the horizontal, it passes through a number of
layers and produces a track in the detector. Such events are easy to identify and reconstruct \cite{Chatterjee:2014vta}. A significant part of ICAL physics program is based on these $\nu_\mu$CC events. In addition, there will be a number of 
events without observable muon tracks \cite{Ajmi:2015qda}. They will consist of (i) charged current events of electron neutrinos
($\nu_e$ and $\bar{\nu}_e$) called $\nu_e$CC events,
(ii) charged current events of tau neutrinos ($\nu_\tau$ and $\bar{\nu}_\tau$) called $\nu_\tau$CC events, (iii) neutral current events of all three types of neutrinos and anti-neutrinos called NC events and (iv) those $\nu_\mu$CC events for which muon track cannot be reconstructed. The last type of events occur due to
low neutrino energy (E$_\nu<$ 0.5 GeV) and/or due to near horizontal direction of neutrino ($|\cos\theta|<$ 0.5 where $\theta$ is the zenith angle of the neutrino).

The magnetised ICAL detector can easily indentify the charge of the muon. The energy and direction of the muon
can be estimated from muon track information \cite{Chatterjee:2014vta}. In a typical $\nu_\mu$CC
event, there are likely to be some hadrons also. The energy of these hadrons can be estimated
using the techniques described in \cite{Devi:2013wxa}. By combining the information from the 
muon and the hadrons, it is possible to obtain an estimate of the neutrino energy \cite{Kaur:2014rxa} and direction.
 
The $\nu_e$CC events produce electrons (positrons) which create a shower in the detector. They lose 
energy very fast and are not able to travel through many layers. The $\nu_\tau$CC 
events \cite{Fukuda:1998mi} are rather small in number because of the large mass of the $\tau$ lepton. The $\tau$ lepton
decays mostly into hadrons and hence these events also look like a shower of hadrons \cite{Migliozzi:2011bj}. 
The only visible part of an NC event consists of hadrons, because the final state neutrino escapes detection.
Hence, the NC events also look like a shower of hadrons.
In the sample of shower-like events, we must include those $\nu_\mu$CC events for which the muon track cannot be reconstructed.

We classify all these
events, without a clear muon track, to be {\it muonless} events. In this paper, we attempt to develop
methods to estimate the energy and direction of the neutrinos which produce such muonless events.

Neutrino interactions are typically described in terms of neutrino energy and direction. The direction 
enables us to calculate the distance the neutrino has travelled \cite{UmaSankar:2006yv}. Thus an estimate of the 
incident atmospheric neutrino energy and direction will allow us to perform a more quantitative 
analysis of any physics topic. 

Using the Nuance neutrino event generator \cite{Casper:2002sd}, we generated 500 years data for ICAL. In generating this data, we assumed
normal hierarchy and used the following values for neutrino parameters:
$\Delta{{m}_{21}}^2 = 7.5 \times 10^{-5}$ eV$^2$, $\Delta{{m}_{31}}^2 (NH) = 2.51\times 10^{-3}$ eV$^2$, $\Delta{{m}_{31}}^2 (IH) = - 2.43\times 10^{-3}$ eV$^2$, $\sin^2\theta_{12}$ =0.31, $\sin^2{2\theta_{13}}$ =0.09, $\sin^2{\theta_{23}}$=0.5  and $\delta_{CP}$=0.
The generated events are then simulated in the ICAL detector using GEANT4 \cite{Redij}.

\section{Incident Energy of the neutrino }

 A more energetic neutrino is expected to give more number of hits in the detector, distributed among more number of layers compared to a less energetic neutrino. For a given neutrino energy, an event with neutrino travelling in vertical direction will have hits in more layers compared to one travelling close to horizontal direction \cite{Hasert:1973ff}. Also, for a given neutrino energy, the NC events have less number of hits compared to CC events. Therefore, the two quantities, 
 \begin{itemize}
  \item number of hits,
  \item number of layers having one or more hits,
 \end{itemize}
 are the basic kinematical variables to be used in the determination of the neutrino energy.
From these we can define additional variables such as average hits per layer (hpl), hit distribution, etc.
We use combinations of these variables to obtain an estimate of the neutrino energy and direction.
  
 We found that there is negligible correlation between the hits and the neutrino energy, or the layers and the neutrino energy. 

\begin{figure}[H]	
 \centering
 \setlength\fboxsep{0pt}
 {\includegraphics[ width=0.9\textwidth]{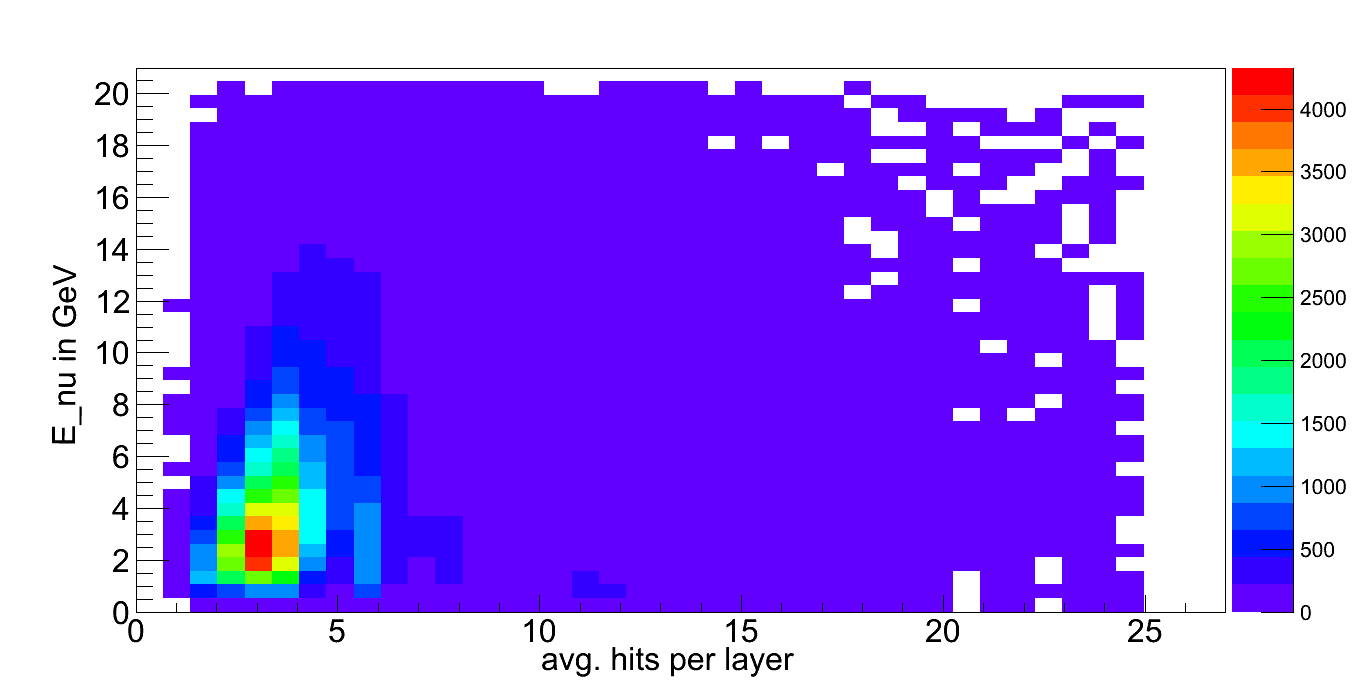} }
 \caption{\small  {Hits/Layers vs. Energy:} Correlation between the incident energy of the neutrino (here $\nu_e$CC shown) and the average number of hits per layer in the event, after the cut \#hits$>$10.  }
\label{Fig1}
 \end{figure}

\begin{figure}[H]	
 \centering
 \setlength\fboxsep{0pt}
 {\includegraphics[ width=0.9\textwidth]{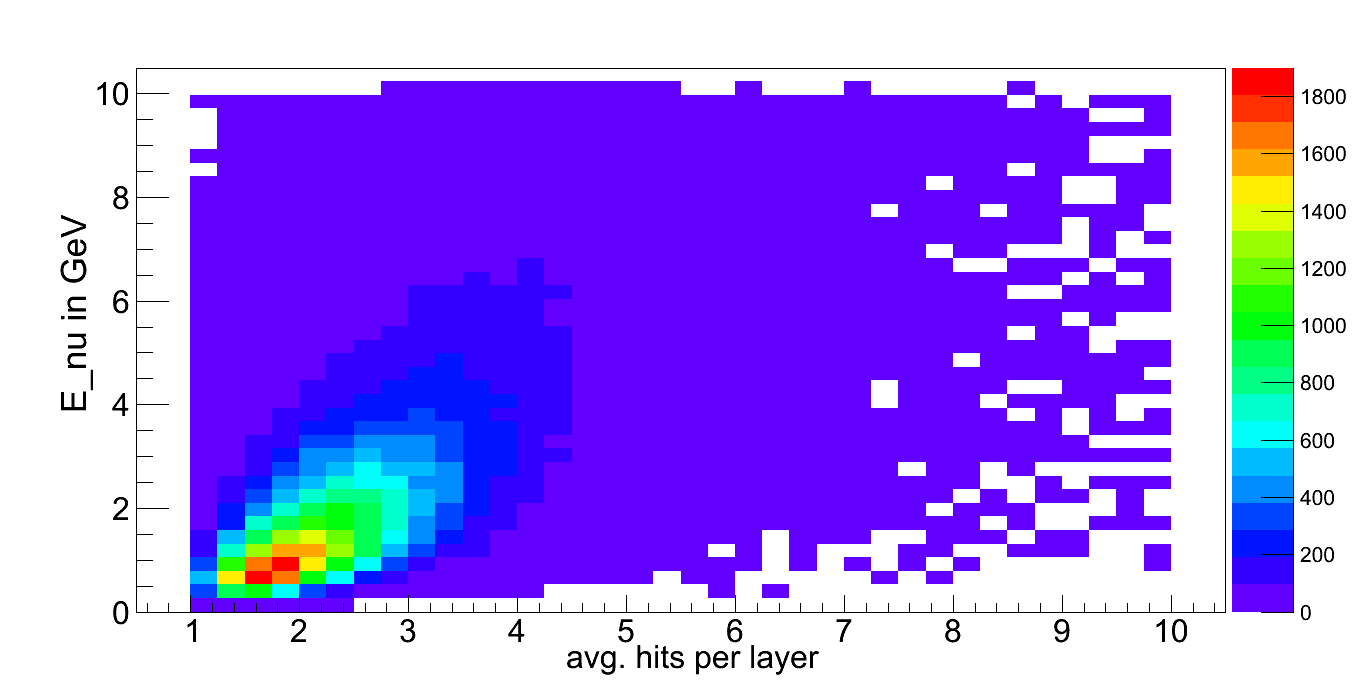} }
 \caption{\small Average hits per layer vs. energy, i.e., dependence of number of hits on the neutrino energy but in a particular layer only, here L=4.   }
\label{Fig2}
\end{figure}

 The variable `hpl' does have some dependence on the neutrino energy but the correlation is too small for it to be an effective parameter for neutrino energy determination. This is depicted in figure~\ref{Fig1}, where we have considered all events which have number of hits greater than 10. 
We expect muonless events to have hits only in a few layers. Therefore,
 we further classify events according to the number of layers in which the hits are observed. For example, events with hits
 only in one layer, events with hits only in two layers etc. upto events with hits only in five layers. A majority of muonless events have hits in five layers or less. 
 If two neutrino interactions of different energies give hits in same number of layers, then the more energetic neutrino should give more hits than the less energetic neutrino. That is, there should be a correlation between neutrino energy and hpl, if the number of layers is held fixed. This is illustrated in figure~\ref{Fig2}, for the case of hits in four layers.

 The  trends observed in figure~\ref{Fig2} for $\nu_e$CC events are also evident in the case of the NC events and those
 $\nu_\mu$CC events which are confined to a few layers. 

\subsection{Calibration of neutrino Energy}

\begin{figure}[H]	
 \centering
 \setlength\fboxsep{0pt}
 {\includegraphics[width=1.\textwidth]{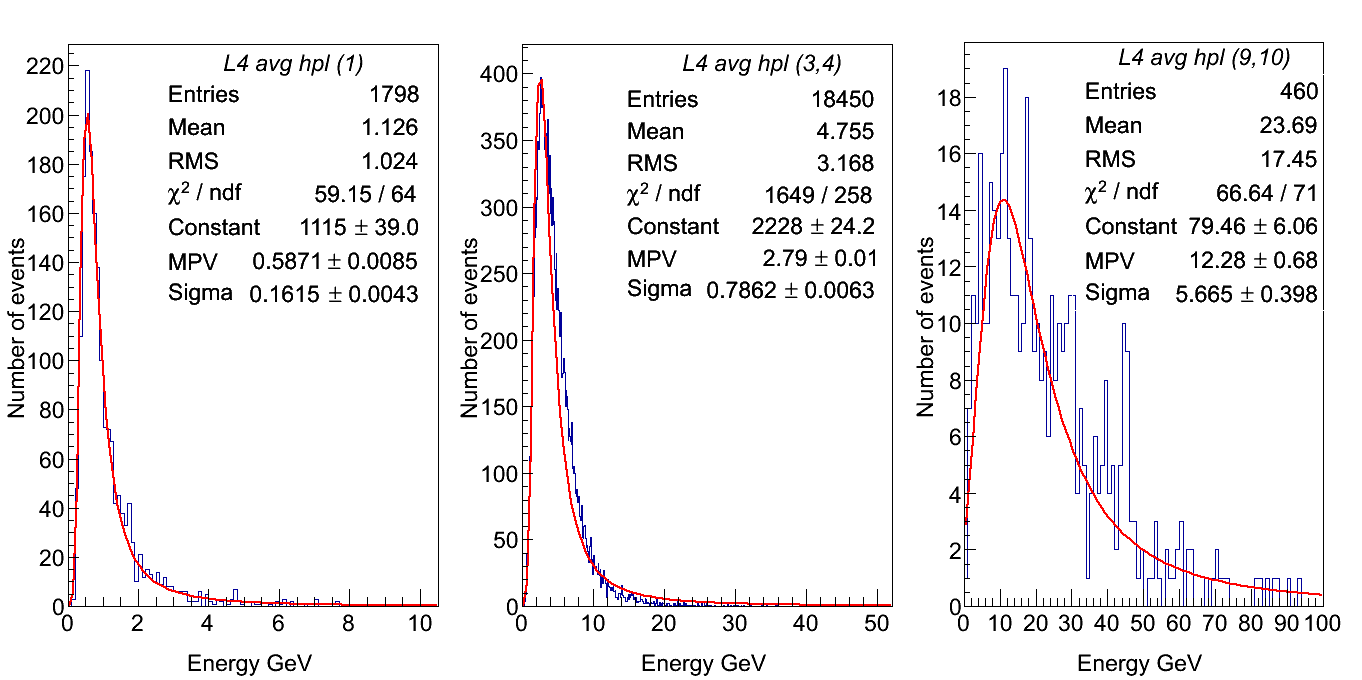} }
 \caption{Neutrino spectra in different bins of hits per layer (from left): (1), (3,4) and (9,10), for events giving hits hits in exactly 4 layers (L=4). The spectra are fitted with Landau distribution function.  }
\label{Fig3}
\end{figure}

As mentioned before, we classify events according to the number of layers containing hits. We have considered values of L (number of layers with one or more hits) =\{2,...,5\}. For a given L, we further subdivide the events in bins of hpl (hits per layer) = \{(1), (1-2), (2-3), ..., (9-10), ...\}. In each of these bins, we have plotted the neutrino spectrum, with the neutrino energy taken from Nuance. A sample of these plots is shown in figure~\ref{Fig3}, for L=4. These spectra show a gradual shift in the peak towards the right along the energy axis, with increasing value of hits per layer. Hence, it appears to provide a reasonable calibration of the neutrino energy. One requires a distribution function to represent each of these spectra. A number of fitting functions have been attempted. Landau distribution is certainly the more obvious one, because it is related to energy loss of charged particles in any medium \cite{Landau:1944if,Wilkinson:1996he}.

 
We describe the Landau distribution by its most probable value (MPV) and its sigma ($\sigma$)
\footnote{The correction -0.22 to the peak position which is recommended for root-fitting at 
lower energies is neglected here since the error bar covers it \cite{Brun:2000es}.}. In our fitting procedure, we have imposed certain conditions. A fit is attempted only for those distributions which have 
at least a few hundred events. In addition, a fit is accepted only if the $\chi^2$/ndf is $\leq$10. 

Other distribution functions like the Vavilov and a few non-standard functional forms have also been attempted. The Vavilov distribution, which is the more general form of the Landau function, fits the spectra well in the bins with moderate values of hits per layer. In fact, the fit is slightly better than the Landau distribution. But the fit to Vavilov distribution is very sensitive to the limits on the fit parameters. 
 Moreover, at very low values of hits per layer and at very high values, Landau distribution 
 gives a much better fit. 
 Here, we consider the calibration of energy over a very wide range of hits per layer. 
 So, we prefer to use one common distribution function over the entire range. 

  \begin{figure}[H]	
 \centering
 \setlength\fboxsep{0pt}
 {\includegraphics[width=1.\textwidth]{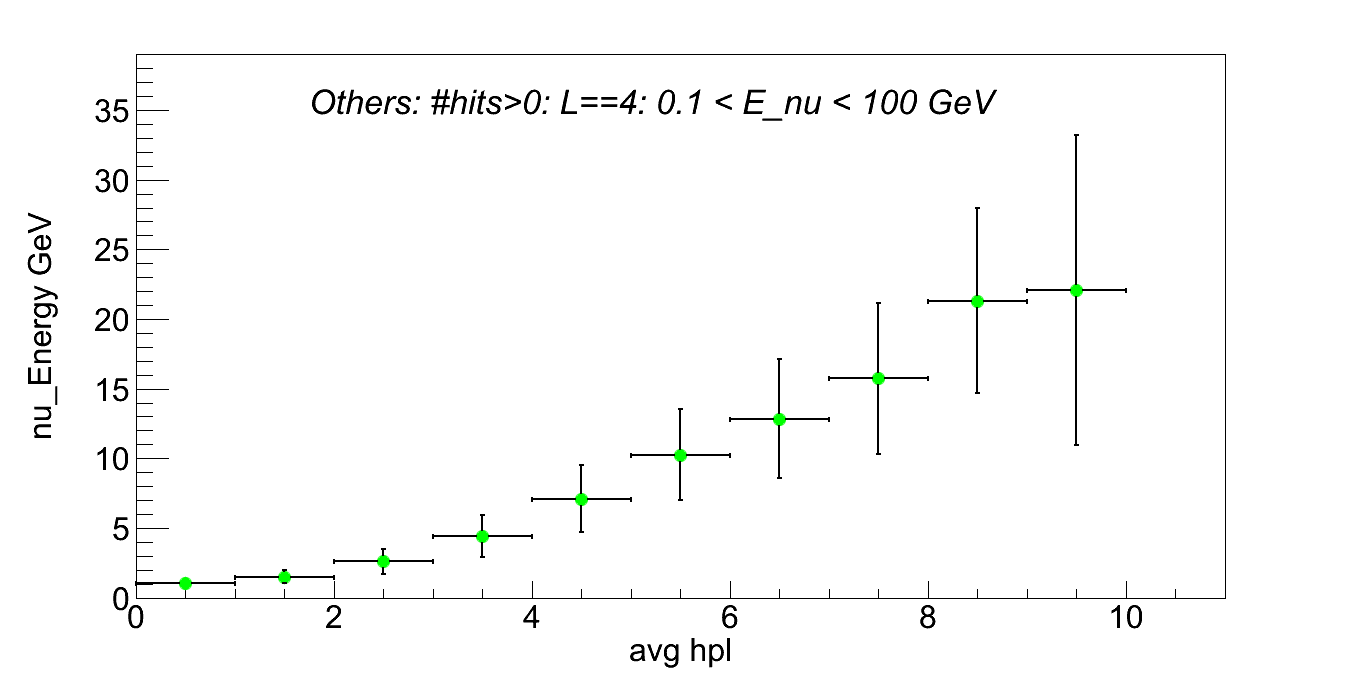} }
 \caption{\small Calibration of $\nu$-Energy vs. Average hits per layer for L=4, for the NC (+$\nu_\tau$CC) events. (Points representation.) The points are given by the Landau peak positions and the `error bars' by Landau $\sigma$ in vertical scale (the horizontal bars cover the hpl bin-width).}
 \label{Fig4}
\end{figure}

 The nature of the energy deposition is different in each of the three types of neutrino events considered here. 
 They are grouped under the  generic name ``muonless'' events but each type has its characteristic interaction properties.
 The NC events contain outgoing neutrinos which give no hits. 
 The electrons present in the $\nu_e$CC events give more hits, in addition to those by the hadrons. 
 The $\nu_\mu$CC events contain those muons which do not give identifiable tracks. Since the energy loss
 in iron is the smallest for muons, $\nu_\mu$CC events are likely to have more number of hits in comparison to
 $\nu_e$CC events of the same energy. 

          	\begin{figure}[H]	
 \centering
 \setlength\fboxsep{0pt}
 {\includegraphics[width=1.\textwidth]{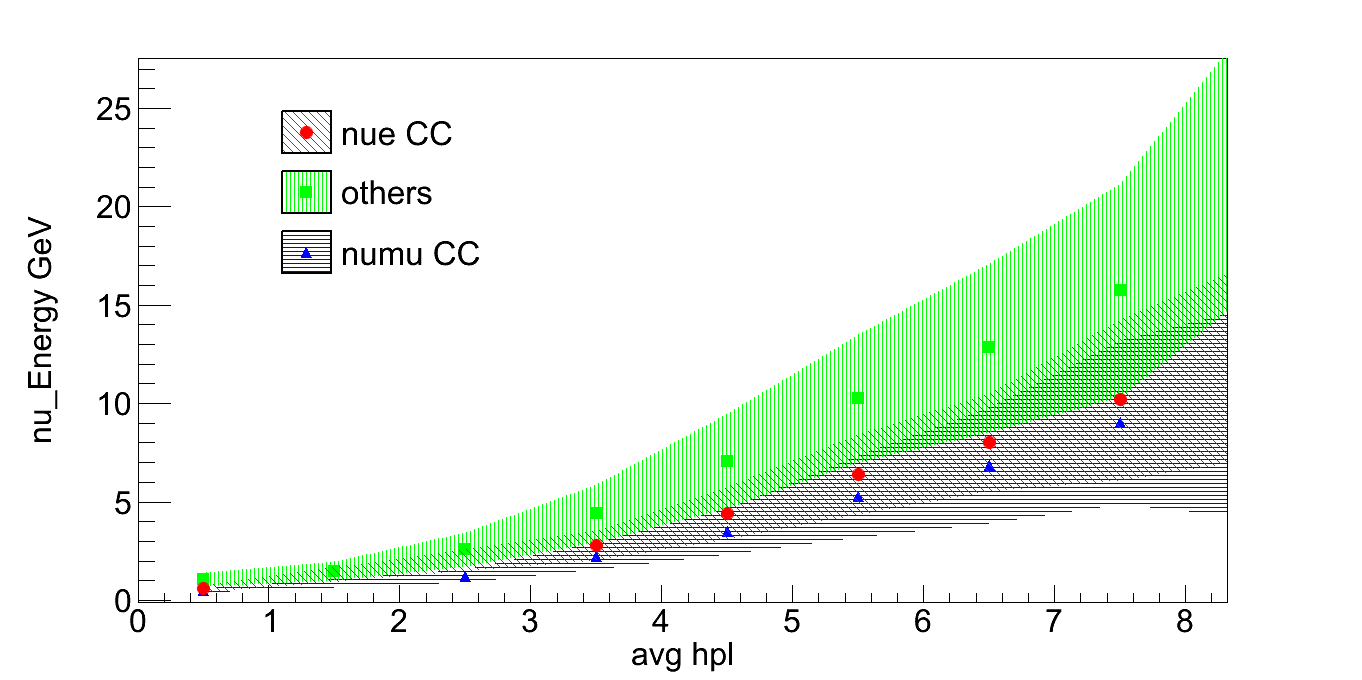}}
 \caption{Approximate Energy calibration of the neutrinos having hits in exactly 4 layers,  to visualise all the three types of muon-less neutrino events ($\nu_e$CC in red, NC in green and $\nu_\mu$CC in blue) all on a uniform scale of hits per layer. }
 \label{Fig5}
\end{figure}
\begin{figure}[H]	
 \centering
 \setlength\fboxsep{0pt}
 {\includegraphics[width=1.\textwidth]{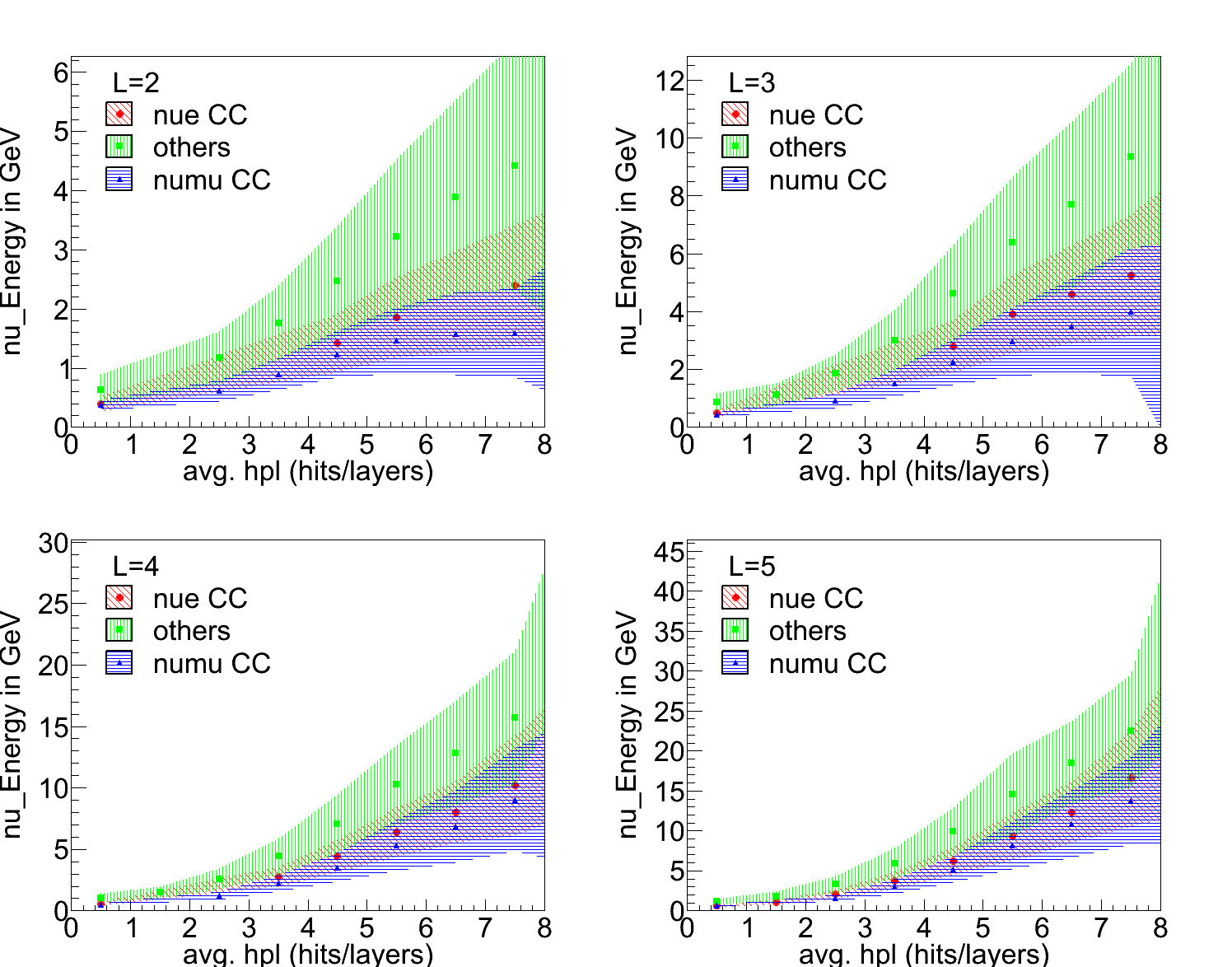} }
 \caption{Approximate Energy calibration of the neutrinos having hits in exactly (from top) 2, 3, 4 and 5 layers. The three types of muonless neutrino events are plotted in different colours. The $\nu_e$CC is in red, NC in green and $\nu_\mu$CC in blue. Note that the scale on X-axis is the same for all plots, but the scale on Y-axis (Energy) increases with increase in L.}
 \label{Fig6}
\end{figure}

The correlation between the
neutrino energy and the hpl is shown in figure~\ref{Fig4}, for NC events with hits in exactly four layers.
The central points are the MPVs of the corresponding Landau distributions and the error bars are $\pm \sigma$s of those
distributions.
Figure~\ref{Fig5} shows this correlation for $\nu_e$CC and $\nu_\mu$CC events also in addition to NC events.
 The NC events as expected give less hits than the other two event types. A chart of energy correlation is prepared in figure~\ref{Fig6}, with ``number of layers'' counting from 2 to 5. 
 The curve in green is more relevant if we are dealing with an events sample rich in NC events. The curve in red is to
 be referred to if we have an events sample rich in $\nu_e$CC events \cite{Ajmi:2015qda}. We have verified that
 a sample of pure $\nu_e$CC events and a sample of events rich in $\nu_e$CC events (with $\sim 60\%$ purity)
 both obey the same correlation plot. The curve in blue focuses on the $\nu_\mu$CC events. It is worth noting that the energy scale increases with increasing L.

\begin{figure}[H]	
 \centering
 \setlength\fboxsep{0pt}
 {\includegraphics[width=1.\textwidth]{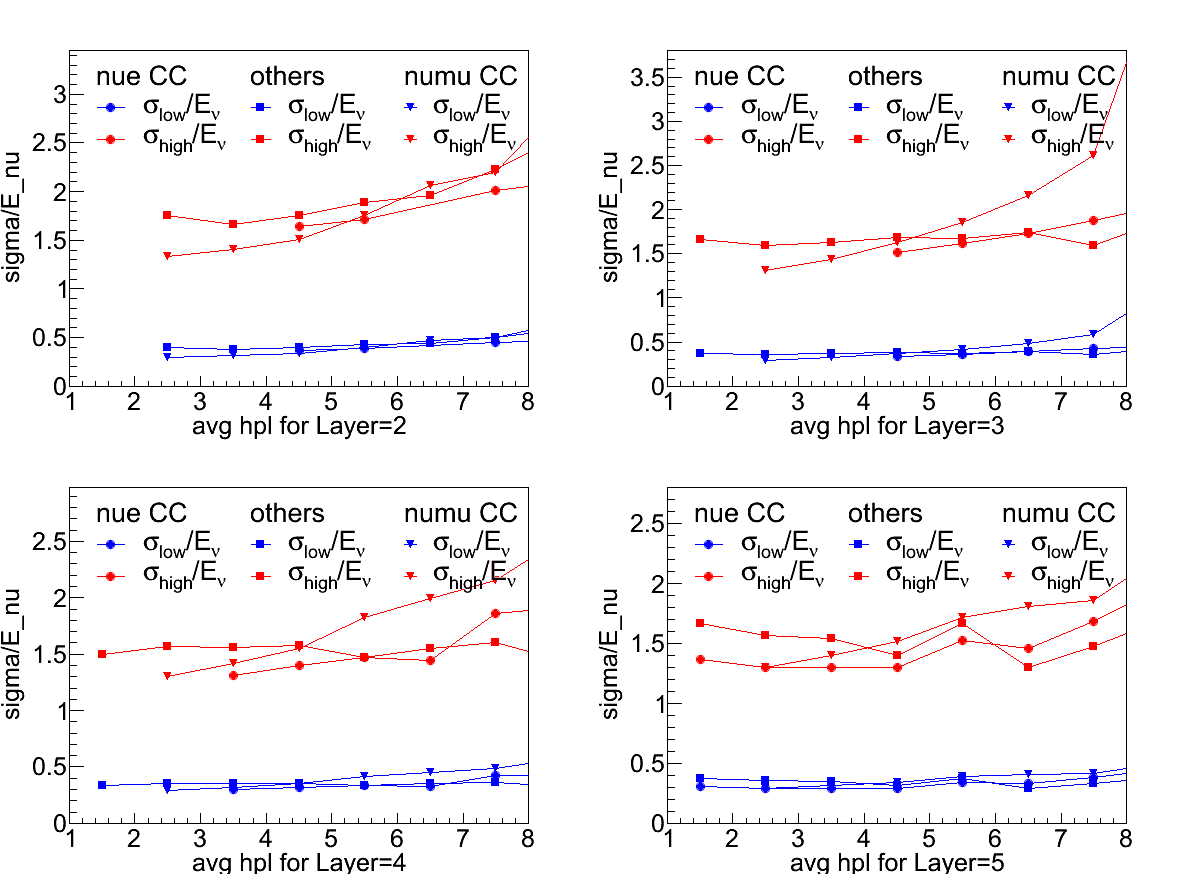} }
 \caption{Energy Resolution: Variation of $\sigma_{\rm low}/{\rm E_\nu}$ and $\sigma_{\rm high}/{\rm E_\nu}$ with hpl, for events giving hits in L = 2, 3, 4 and 5 layers. [E$_\nu$ refers to the MPV of the Landau distribution.] The resolution plots for all three event types are shown.}
 \label{Fig7}
\end{figure}

For Landau distribution, $68\%$ of the events fall within the range  
(MPV - 1.02 $\sigma$, MPV + 4.65 $\sigma$) \cite{elossppt}. This leads to
the definitions $\sigma_{\rm low} = 1.02 \sigma$ and $\sigma_{\rm high} = 4.65 \sigma$.
The plots of $\sigma_{\rm low}/{\rm MPV}$ and $\sigma_{\rm high}/{\rm MPV}$
are shown in figure~\ref{Fig7} for $L = 4$. The lower values are about $0.3$ whereas
the higher values are about $1.5$ for all the cases. 
For given values of L and hpl, we can assign the event the most probable value of the neutrino energy. 
We can also estimate the probability of the event having a given energy
based on the MPV and $\sigma$ of the corresponding Landau distribution.

\section{Direction of the neutrino}

 The direction of a neutrino is given in terms of the polar angle $\theta$ and the 
 azimuthal angle $\phi$. The neutrino flux is expected to be symmetric in $\phi$,
 except for a small east-west asymmetry arising due to the earth's magnetic field.
 The polar angle of neutrino is certainly of greater importance because it determines
 the distance travelled by the neutrino, given by the expression
 $
  L=\sqrt{(R+h)^2 - (R \sin\theta)^2}-R \cos\theta,
$ where $R$ is the radius of the earth and $h$ is the atmospheric height at which the neutrino
is produced \cite{UmaSankar:2006yv}.
 We see from the above equation that $\cos\theta$=1 (-1) for the vertically up (down) going neutrinos.
 To do oscillation physics, we need to distinguish between upgoing neutrinos which
 travel thousands of km and down going neutrinos which travel only tens of km. In
 addition, it will be useful to distinguish between neutrino events in horizontal and vertical
 directions.
 
 In muonless events, there are no tracks to serve as an easy handle in determining the neutrino direction.
 Since these events have hits in five layers or less, the timing information is not useful in 
 distinguishing between upward going and downward going events.
 Here we attempt to develop a method to find a correlation between the hit pattern and the neutrino
 direction. The following methods are tested with the $\nu_e$CC events and are found to work. 
 Both NC events and $\nu_\mu$CC events without muon tracks (those for which hits are limited to five layers or less)
 also behave in a similar fashion. Minimal selection criteria may be used to select such event sample \cite{Ajmi:2015qda}.

We have checked that the neutrino direction is not particularly correlated to the number of hits 
or the number of layers hit. Hence, devising criteria based on hits and layers or even hits per layer is not 
effective in finding a way to recognize the angular information of an incident neutrino. 
 
 \subsection{Algorithms devised for neutrino angle depiction}
 We have considered the problem in two parts: Firstly, we try to distinguish between the vertical to near-vertical 
 events from the horizontal to near horizontal neutrino events. Secondly, we attempt to tell apart the up-going 
 neutrinos from the down-going neutrinos. We have devised a number of algorithms and tested their efficacy in 
 determining the $\nu_e$ direction.
 
 \subsubsection{Horizontal or Vertical Direction}
  
  The vertical or near vertical events should have shorter average horizontal spread than the horizontal or near 
  horizontal events. Having taken this cue, the maximum total spread of a neutrino event in horizontal plane is 
  studied. This parameter is defined as follows: 
   A hit point in ICAL detector refers to the pair of (x,y) coordinates of the hit in an RPC. 
   The layer number of the RPC gives the z coordinate. The signals from the detector are read as the strip numbers 
   along the X or Y direction. Hence, the hit points are devised by making all possible combinations of the signal giving X-Y strip numbers. Now, the distance between any two points in a given layer is given by
  $ D = \sqrt{((x_2-x_1)^2+(y_2-y_1)^2}$), for a pair of hits in a layer.
 
  We can also define the horizontal distance between any two hit points by a similar formula,
  even if the hits are in different layers. This horizontal distance is the projection, of the
  total distance between the two points, on the horizontal plane.
  The maximum horizontal distance, the variable we use here, is defined to be the maximum value
  of the horizontal distance of an event. It is the maximum of projected lengths, on the horizontal
  plane, of the distances between any two hits in the event. 
  
      	\begin{figure}[H]	
 \centering
 \setlength\fboxsep{0pt}
 {\includegraphics[width=1.\textwidth]{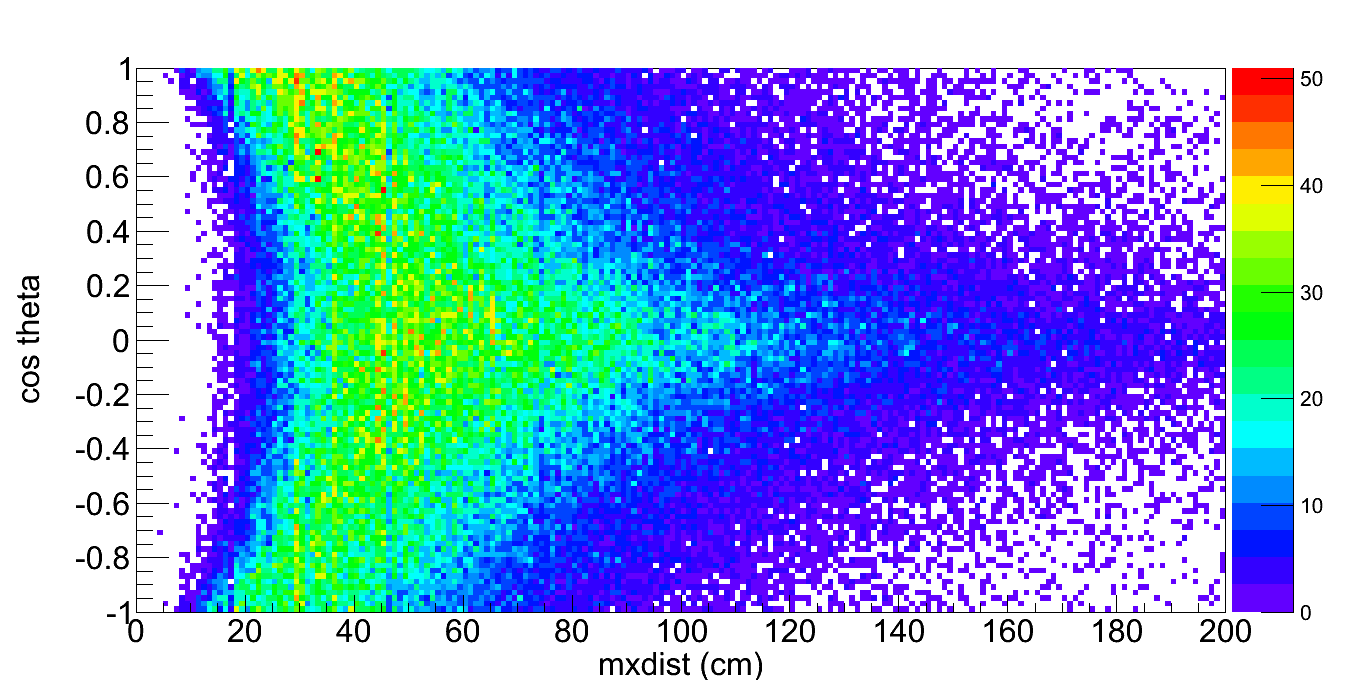}} 
 \caption{Correlation between the cosine of incident theta of the neutrino (here $\nu_e$ shown) and the maximum horizontal spread in an event for the 500years NH data. E$_\nu$=\{0.8,20\}GeV. }
 \label{Fig8}
\end{figure}

From figure~\ref{Fig8}, we see that there is a small correlation between direction of the neutrino (vertical
or horizontal) and the maximum horizontal distance. The neutrinos incident in the vertical cone have 
less horizontal spread than the ones in the horizontal cone. For example, an upper cut of 30 cm on
the maximum horizontal distance, gives above 75\% vertical events ($|\cos \theta| > 0.4$ or those within a 
cone of angle $65^\circ$ about the vertical direction) in the selected sample.

\subsubsection{Up-going or Down-going neutrinos}
The muonless events do not travel through many layers, unlike an event with a muon track. So, one needs to find a method/way to tell apart the upgoing neutrinos from the downgoing ones, in order to extract better physics information.

\begin{figure}[H]	
 \centering
 \setlength\fboxsep{0pt}
 {\includegraphics[width=1.\textwidth]{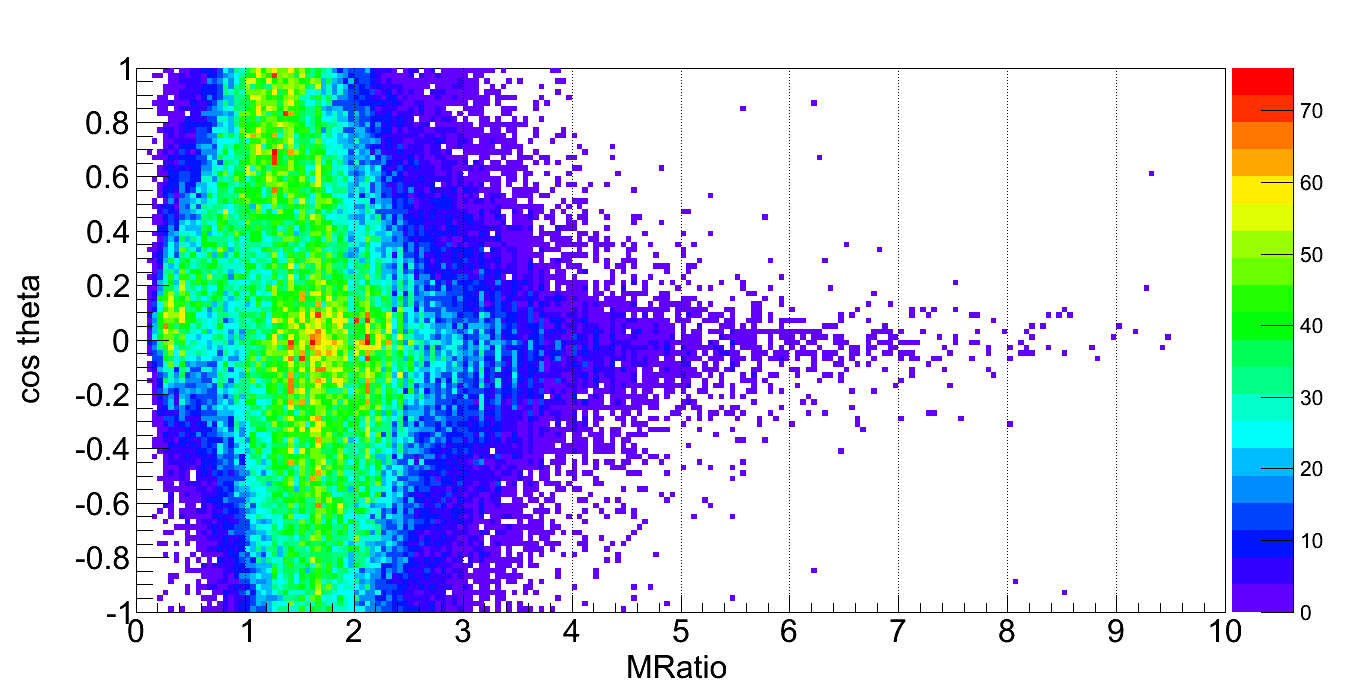} }
 \caption{Correlation between the cosine of incident theta of the neutrino (here $\nu_e$ shown) and the ratio of layer-hits mean and rms in an event for the 500years NH data. E$_\nu$=\{0.8,20\} GeV. }
 \label{Fig9}
\end{figure}

Finally, an algorithm using the hit distribution pattern can select upgoing neutrinos to an efficiency of 70\%, 
and is described as follows.
Assuming the number of hits across the layers as a type of distribution, their mean and the standard deviation from the mean (rms) are calculated. The details of this calculation are explained in \cite{Ajmi:2015qda}. 
These values of the mean and the rms of the layerhits distribution, hardly show any dependence on the direction of the neutrino.
But the ratio of layer-hits mean to layer-hits rms (``MRratio'') shows a dependence on whether a neutrino is upgoing or downgoing, as seen in figure~\ref{Fig9}.
The figure clearly shows that the lower values of this variable called MRratio selects mostly upgoing events and vice versa.

\subsection{Estimation of the neutrino cos(theta): }

\begin{figure}[H]	
 \centering
 \setlength\fboxsep{0pt}
 {\includegraphics[height=.98\textheight]{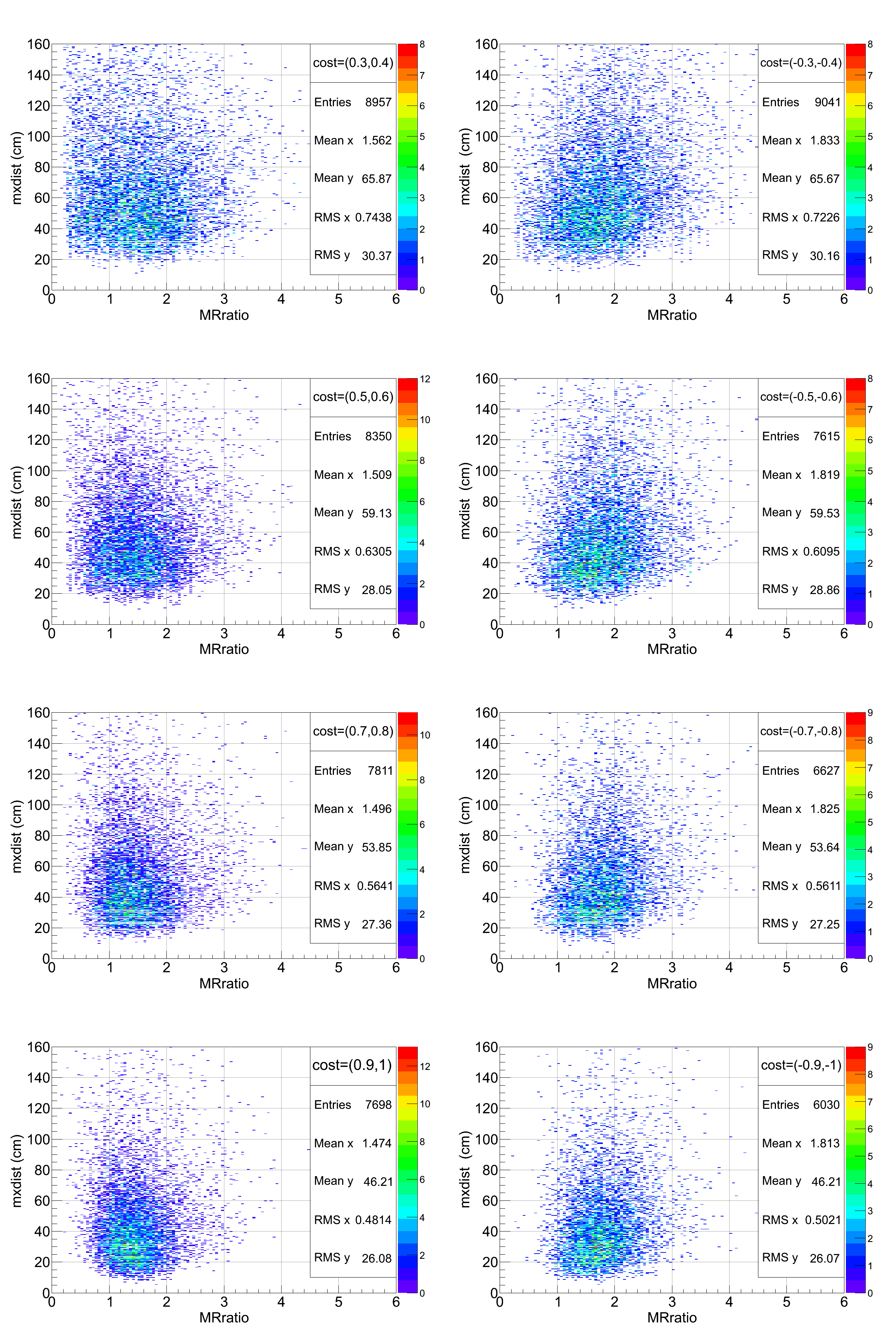} }
 \vskip -4ex
 \caption{\small Correlation of MRratio and the mxdist, for $\nu_e$CC events in bins of cos theta (here only some of them are shown), for the 500 years NH data. E$_\nu$=\{0.8,20\} GeV. }
 \label{Fig10}
\end{figure}

 The ratio of layer-hits mean to rms (``MRratio'') gives a way to distinguish the upgoing neutrino events 
 from the downgoing ones. The maximum horizontal distance (``mxdist'') provides us with the ability to separate 
 vertical events from the horizontal events. 
 We have attempted to find some correlation between the two of them. 2-D histograms of these variables show gradual shifts in the peak positions of such distributions, in varying bins of neutrino direction. Figure~\ref{Fig10} shows some of them. The 2D projection of the surface plot, of each of these distributions, show them to be symmetric along the MRratio-axis and asymmetric along the mxdist-axis,  with a tail towards higher values of mxdist.
 
 The left hand side panel in figure~\ref{Fig10} shows distributions of the upgoing neutrinos in a few selected bins of cos$\theta$. The 
right hand side panel shows the same for the down going neutrinos. 
The comet-like distribution (tail more notably) has shifted from left (in figures of the left hand side panel) 
to the right (in figures on the right hand side panel), along X-axis, i.e. in the direction of increasing MRratio.
The  head of the ``comet''  moves downwards (towards smaller values of mxdist), as one goes from the top to the bottom of these set of panels (as cos$\theta$ varies from vertical to horizontal) in figure~\ref{Fig10}.
 

The directional information conveyed by these two variables independently, may be combined to obtain the resultant neutrino direction.
Hence, one can present a calibration of cos$\theta$ vs. MRratio and mxdist in 3-dimensions.

\begin{figure}[H]	
 \centering
 \setlength\fboxsep{0pt}
{\includegraphics[width=1.\textwidth]{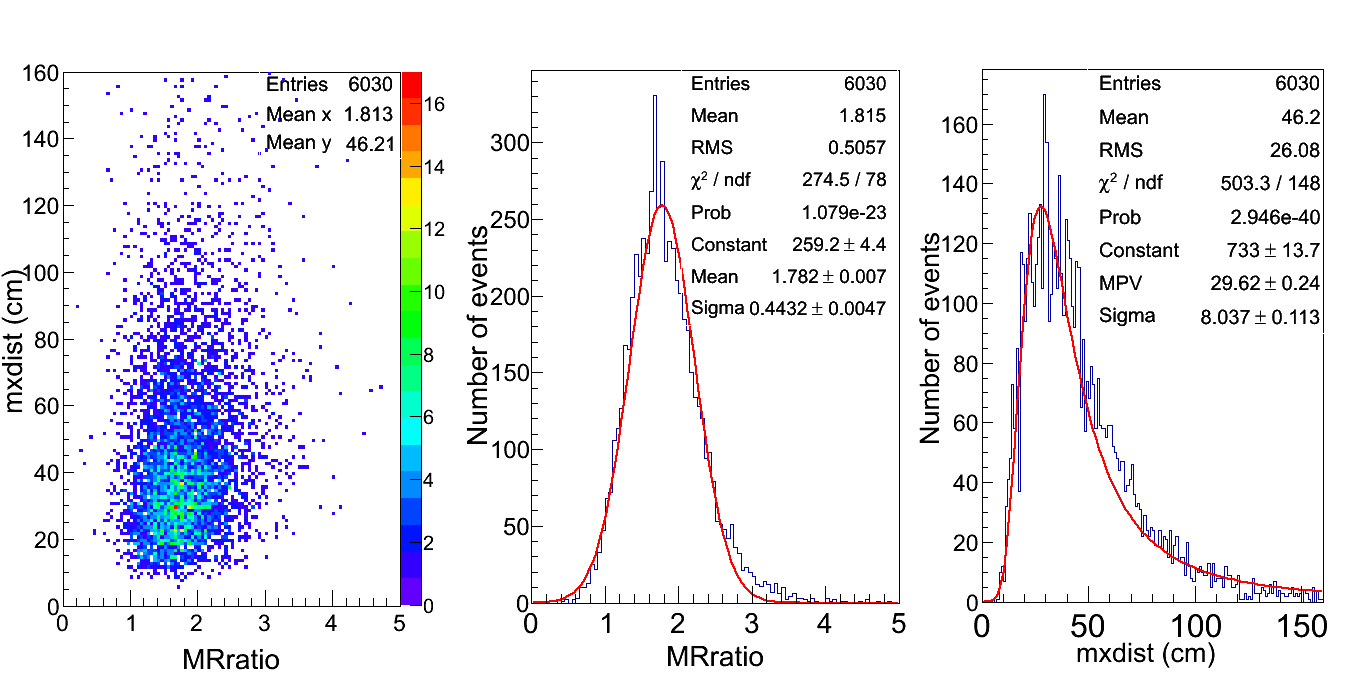} }
 \caption{Left: 2D projection of the correlation of the MRratio and mxdist; middle: Fitting a gauss function on the XZ projection of the plot; Right: Fitting a landau function on the YZ projection of the plot.}
 \label{Fig11}
\end{figure}

The 2-Dimensional histograms can be fitted with appropriate surface distribution functions. The peak of such a fitted function gives us the coordinates for the 3D-calibration plot. The errors can be quoted from the sigma of those peaks. 
Looking at the comet-like distribution from figure~\ref{Fig10}, one may propose a Gaussian distribution fit along X-axis (MRratio), and a Landau distribution fit along Y-axis (mxdist). Figure~\ref{Fig11} shows such an example.

\begin{figure}[H]	
 \centering
 \setlength\fboxsep{0pt}
{\includegraphics[width=1.\textwidth]{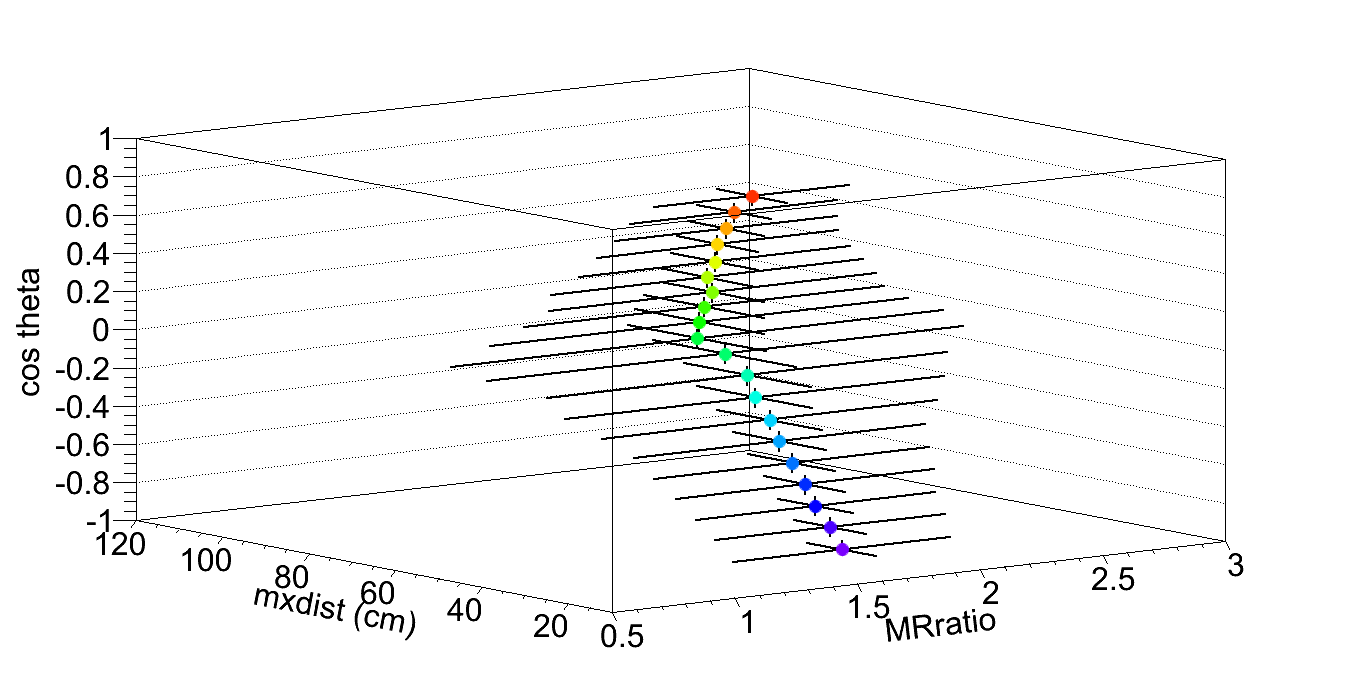}}
 \caption{\small {\bf ``The Skewed-Hair-pin Structure'': }Calibration of cos$\theta$ with respect to the plane spanned by the layer-hits mean to rms ratio and maximum spread,   for the 500years NH data (here $\nu_e$ shown). E$_\nu$=\{0.8,20\}GeV.  }
 \label{Fig12}
\end{figure}

The angular estimation can thus be done in terms of a 3D-calibration plot of MRratio along X-axis, mxdist along Y-axis and costheta along Z-axis, as in figure~\ref{Fig12}. 
The X-axis contains the Gaussian mean of the MRratio with $\pm$ Gaussian sigma as the standard deviation, in that costheta bin. The Y-axis contains the Landau peak position of the maximum spread with $\pm$ Landau sigma as the standard deviation. 
The costheta is along Z-axis with binwidth of 0.05. 

     	\begin{figure}[H]	
 \centering
 \setlength\fboxsep{0pt}
{\includegraphics[width=1.\textwidth]{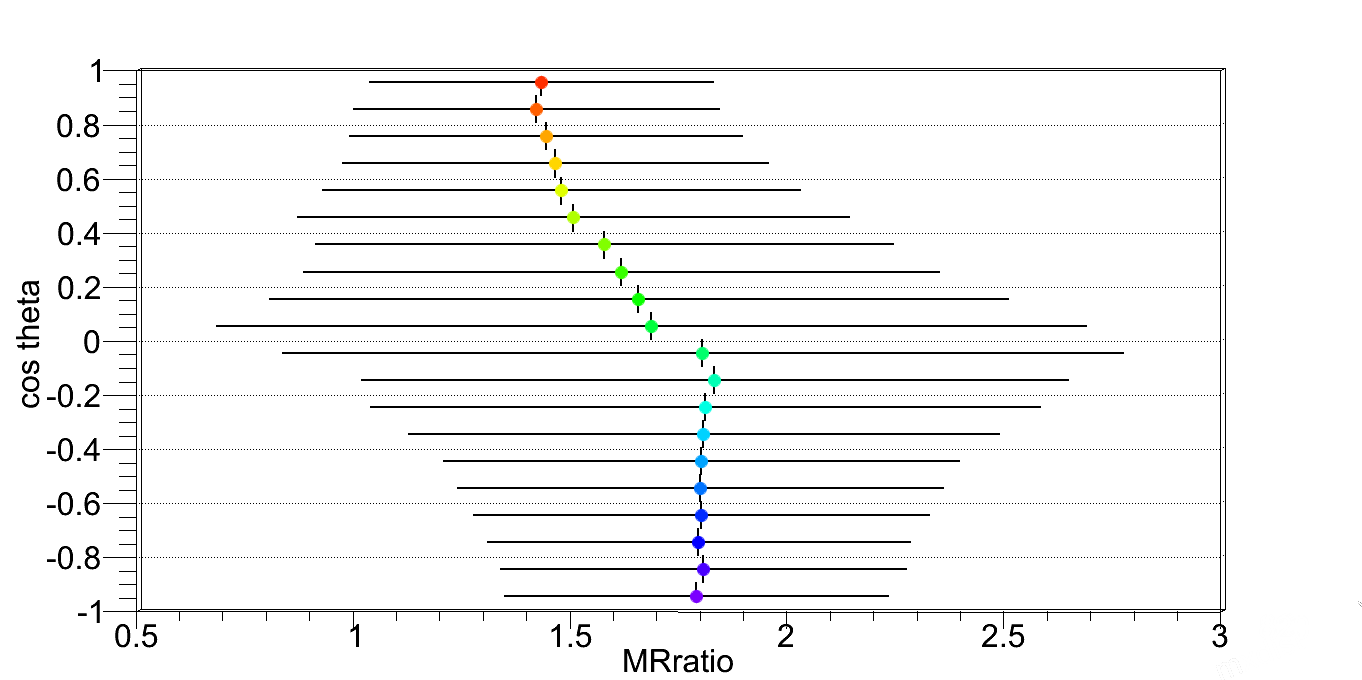}}
 \caption{\small The X-Z projection of the figure~12, i.e. gaussian fitting of the distribution. }
 \label{Fig13}
\end{figure}

     	\begin{figure}[H]	
 \centering
 \setlength\fboxsep{0pt}
{\includegraphics[width=1.\textwidth]{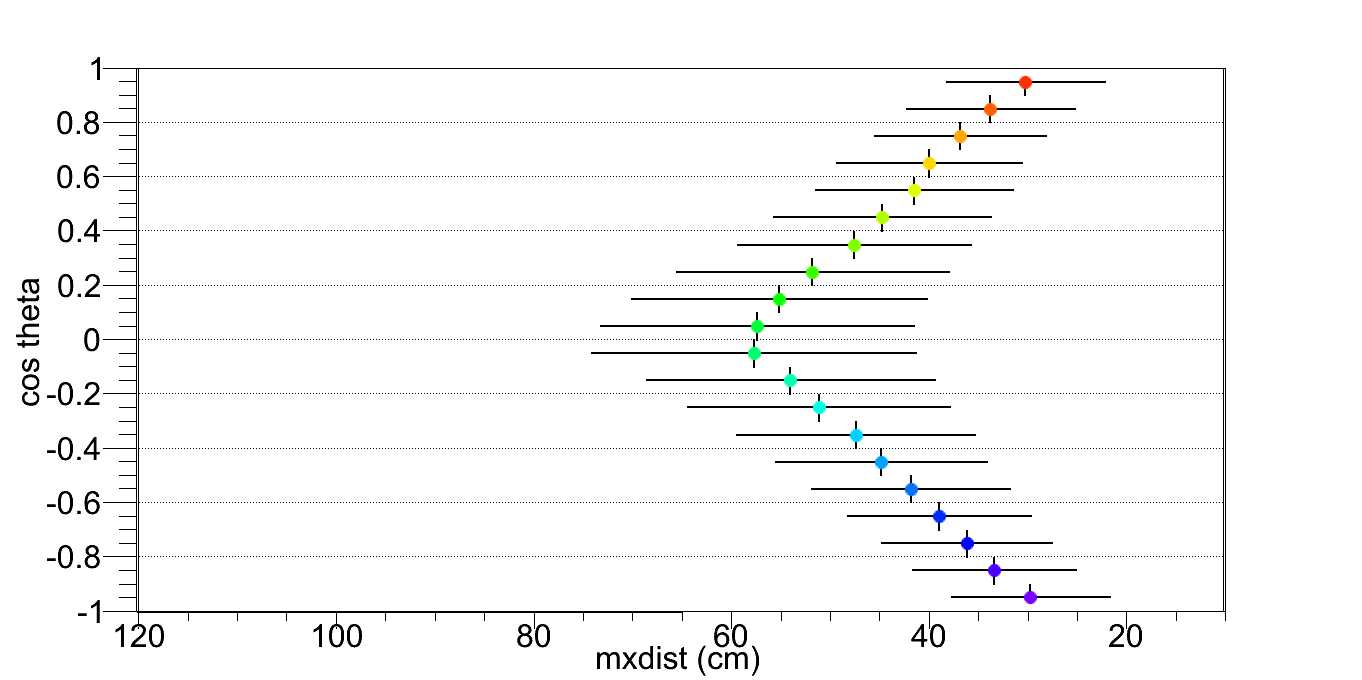}}
 \caption{\small The Y-Z projection of the figure~12, i.e. Landau fitting of the distribution.  }
 \label{Fig14}
\end{figure}
 
 The XZ and YZ projections have been shown in figure~\ref{Fig13} and \ref{Fig14} to help in visualising figure~\ref{Fig12}. But the estimation 
 of the angle is preferably made from the 3D-Calibration Plot in figure~\ref{Fig12}. As we see from figure~\ref{Fig14}, mxdist 
 gives a reasonable estimate of the modulus of cos$\theta$. Then we can use MRratio to break the degeneracy between
 the up and down going events. 
 

\section{Summary and Conclusions}
An effective method has been devised to estimate the energies of the neutrino events which give no clear muon track. 
The energy of the incident neutrino can be well estimated by following the approximate calibration curves in figure~\ref{Fig6}.
If the event sample contains equal proportions of $\nu_e$CC, NC and $\nu_\mu$CC events, then the uncertainty in the 
estimated energy is rather large. However, it is possible to choose event samples which are rich in $\nu_e$CC
or in NC events \cite{Ajmi:2015qda}. For those muonless samples, it is possible to get a good energy estimate. 

One of the important types of such muonless events is the NC interaction.  It must be remembered that NC events have an outgoing neutrino, which does not leave any signature. 
So, this method serves as an eventual solution to estimating the energy of such incident neutrinos. 

The energies of the neutrinos in case of the muon-track containing events can be determined from the muon track informations added to that from the hadron energy calibration. 
Now, with the present method, one can estimate the energies of neutrinos that do not give muon tracks, i.e., the Neutral Current interactions, $\nu_e$CC interactions as well as $\nu_\mu$CC events which lack identifiable muons. 
{\bf Hence, to sum up, we can estimate the energies of all kinds of neutrinos (active) that are detected by the INO-ICAL Detector. }

We can even make a crude estimation of the angle or direction of the incident neutrino for muonless events. The two variables, mxdist and MRratio have been defined to get this estimation. 
Mxdist can distinguish the vertical/near-vertical events from the horizontal/near-horizontal events, which leads to a degeneracy between
upgoing and downgoing events. MRratio removes this degeneracy and makes it possible to get an approximate estimate
of the neutrino direction.

 \section{Acknowledgement}
 We express our deep gratitude to all our co-members of the INO Collaboration. We are thankful for the suggestions from Prof. Gobinda Majumder and discussions with our fellow members during the meetings have gone a great way in guiding us rightly through the project. 
  We are also grateful to the Department of Atomic Energy (DAE) for financial support.

 \bibliographystyle{apsrev4-1}

 \bibliography{/home/ali/UmaS/reports/Aug2014/SML_bibtex_inspire}

 \end{document}